\documentclass{article}
\usepackage{amsmath}
\usepackage{amssymb}
\usepackage{amsfonts}

\input{tcilatex}
\begin{document}

\title{The Art of Probability Assignment\thanks{%
Presented at MaxEnt 2012, the 32nd International Workshop on Bayesian
Inference and Maximum Entropy Methods in Science and Engineering, July
15-20, 2012, Garching near Munich, Germany.}}
\author{Vesselin I. Dimitrov \\
Idaho Accelerator Center, Idaho State University\\
Pocatello ID 83209, USA\\
}
\maketitle

\begin{abstract}
The problem of assigning probabilities when little is known is analized in
the case where the quanities of interest are physical observables, i.e. can
be measured and their values expressed by numbers. It is pointed out that
the assignment of probabilities based on observation is a process of
inference, involving the use of Bayes' theorem and the choice of a
probability prior. When a lot of data is available, the resulting
probability are remarkably insensitive to the form of the prior. In the
oposite case of scarse data, it is suggested that the probabilities are
assigned such that they are the least sensitive to specific variations of
the probability prior. In the continuous case this results in a probability
assignment rule wich calls for minimizing the Fisher information subject to
constraints reflecting all available information. In the discrete case, the
corresponding quantity to be minimized turns out to be a Renyi distance
between the original and the shifted distribution.
\end{abstract}

\section{Introduction}

The problem of probability assignment has been stirring debates and
controversy ever since Laplace introduced the notion of indifference as an
argument in specifying prior distributions. He thus started a quest for a
statistical Holy Grail: prior distributions reflecting ignorance. Today,
more than two centuries later, a satisfactory solution to this problem is
still elusive. In what follows we offer a physicist's take on the somewhat
narrower problem of assigning probabilities for measurable quantities, or,
as physicists call them, observables. Strict space limitations will force
the expos\'{e} to be much more concise than it should have been, but
hopefully the main message will be able to come through.

\section{Probabilities as opinions: an objective take on subjectivity}

When we state "\textit{A }has a probability \textit{p} of being true", what
we really mean is "We don't know whether \textit{A} is true or false, yet we 
\emph{believe} that, if our world existed together with a number of its
replicas, \textit{A} would be true in \textit{pN} out of \textit{N} of them
when \textit{N}$\rightarrow \infty $". Now, it should be evident that,
because of the implied limit procedure, there is no practical way of
verifying this statement. One cannot possibly reproduce a given physical
situation down to its ever minute details several, let alone infinite,
number of times - hence "we believe". Without this leap of faith no rational
science would be possible. An example of this sort of belief can be found in
Mechanics - we know that material points do not exist, but we\emph{\ believe}
that if they did, they would behave according to the Fist, Second and Third
Newton's laws. The source of our faith in this case are countless
observations of the behavior of real objects from afar. It is practice that
sorts out "good" from "bad" beliefs. Different people, however, have
different experiences, so beliefs are subjective and may differ
significantly from one person to another. It is, therefore, of significant
interest to inquire what is it that makes it possible for rational agents to
agree among themselves on what exactly they observe. To that end, let us try
to walk in Laplace's "inverse probabilities" footsteps in analyzing how
opinions are formed from observations. The following builds on [2].

\subsubsection{The Anatomy of a Measurement}

For our purposes, we shall simplistically call "a measurement" a
well-defined procedure to put a real number in correspondence with a
physical phenomenon. Usually we have a good idea what the range $\mathcal{R}%
=[a,b]$ of this number is, but the practicalities of the particular
procedure prevent it from being precise. Thus, instead of a real number $\in
\lbrack a,b]$ the outcome of a single measurement is rather a pointer
(index) $i$ to a subinterval $D_{i}\subset \lbrack a,b]$ where $[a,b]=\cup
_{i=1}^{n}D_{i}$ and $D_{i}\cap D_{j\neq i}=\varnothing $. Repeating%
\footnote{%
"Repeating" here is a misnomer - what is meant is an "ensamble" of replicas
of the world with one measurement performed in each of its members.} the
measurement $m$ times we end up with a histogram of $n$ bins where each bin $%
i$ contains $s_{i}$ - the number of times the measurement fell in that bin.
Obviously, $\sum_{i=1}^{n}s_{i}=m$. Now, if a result in bin $i$ had an
assigned probability $p_{i}$ in a single measurement, probability theory
teaches us that the probability of a set $\{s\}$ is of the multinomial form $%
P(\mathbf{s}|\mathbf{p})=\frac{m!}{\Pi _{i=1}^{n}s_{i}!}%
p_{1}^{s_{1}}p_{2}^{s_{2}}\cdots p_{n}^{s_{n}}$. We, however, are interested
in the opposite situation - the results of the measurements $\{s\}$ are
known, and we want to assign probabilities. In this case we recognize $P(%
\mathbf{s|p})$ as the likelihood and apply the Bayes theorem to obtain the
probability of an assignment $\{p\}$ given the measurements $\{s\}$%
\begin{equation}
P(\mathbf{p}|\mathbf{s})=\mathcal{N}^{-1}\text{ }p_{1}^{s_{1}}p_{2}^{s_{2}}%
\cdots p_{n}^{s_{n}}\pi (\mathbf{p})\delta (\Sigma _{i=1}^{n}p_{i}-1)
\end{equation}%
where $\mathcal{N}$ $=\int d^{n}pp_{1}^{s_{1}}p_{2}^{s_{2}}\cdots
p_{n}^{s_{n}}\pi (\mathbf{p})\delta (\Sigma _{i=1}^{n}p_{i}-1)$ is a
normalization factor. $\pi (\mathbf{p})$ is a probability prior which
originates in whatever knowledge we have about the phenomenon in question,
the measuring procedure and the structure of the domain's decomposition $%
[a,b]=\cup _{i=1}^{n}D_{i}$. For example, one might find it reasonable to
assign prior probability proportional to the measure (length) of $D_{i}$
etc. \bigskip

\subsubsection{The R\^{o}le of the Probability Prior}

With (eqn.1) the most natural way to assign the individual probabilities is
as the expectations%
\begin{equation}
<p_{k}>=\mathcal{N}^{-1}\int d^{n}pp_{1}^{s_{1}}p_{2}^{s_{2}}\cdots
p_{k}^{s_{k}+1}\cdots p_{n}^{s_{n}}\pi (\mathbf{p})\delta (\Sigma
_{i=1}^{n}p_{i}-1)
\end{equation}%
where the integration is over the unit hypercube $p_{.}\in \lbrack 0,1]$.
For a uniform prior $\pi (\mathbf{p})=1$ the integration [1] produces $%
<p_{k}>=\frac{s_{k}+1}{m+n}=\frac{1}{1+\frac{n}{m}}\left( f_{k}+\frac{1}{m}%
\right) $ where $f_{i}=s_{i}/m$ are the "sample frequencies". The variances
of this assignment are easily calculated to be $<(\Delta p_{k})^{2}>=\frac{1%
}{m+n+1}<p_{k}>\left( 1-<p_{k}>\right) $. For a different prior -- uniform
on a quadrant of the hypersphere defined by $p_{i}=\omega _{i}^{2}$ -- the
integrals have been evaluated in [2] as $<p_{k}>=\frac{1}{1+\frac{n}{2m}}%
(f_{k}+\frac{1}{2m})$ and $<(\Delta p_{k})^{2}>=\frac{1}{m+1+n/2}%
<p_{k}>\left( 1-<p_{k}>\right) $. For a general prior we use the average
value theorem from Analysis to obtain 
\begin{eqnarray*}
&<&p_{k}>=\frac{\pi (\varsigma _{k}^{\prime })}{\pi (\mathbf{\varsigma })}%
<p_{k}>_{0} \\
&<&(\Delta p_{k})^{2}>=\frac{\pi (\varsigma _{k}^{\prime \prime })}{\pi (%
\mathbf{\varsigma })}<(\Delta p_{k})^{2}>_{0}+\left[ \frac{\pi (\varsigma
_{k}^{\prime \prime })}{\pi (\mathbf{\varsigma })}-\left( \frac{\pi
(\varsigma _{k}^{\prime })}{\pi (\mathbf{\varsigma })}\right) ^{2}\right]
<p_{k}>^{2}
\end{eqnarray*}%
where $\mathbf{\varsigma ,\varsigma }_{k}^{\prime }$ and $\mathbf{\varsigma }%
_{k}^{\prime \prime }$ are points in the unit hypercube close to the maxima
of $\Pi _{i=1}^{n}p_{i}^{s_{i}}\delta (1-\sum_{i=1}^{n}p_{i}),$ $p_{k}\Pi
_{i=1}^{n}p_{i}^{s_{i}}\delta (1-\sum_{i=1}^{n}p_{i})$ and $p_{k}^{2}\Pi
_{i=1}^{n}p_{i}^{s_{i}}\delta (1-\sum_{i=1}^{n}p_{i})$, correspondingly, and
the zero-subscript quantities are those corresponding to uniform prior.
Assuming \emph{abundance of data} (large $s_{i}$, and, correspondingly, $m$)
and smooth prior, it can be shown that $\mathbf{\varsigma }_{k}^{\prime }%
\mathbf{-\varsigma }\simeq \frac{\mathbf{n}_{k}}{m}$ and $\mathbf{\varsigma }%
_{k}^{\prime \prime }\mathbf{-\varsigma }\simeq \frac{2\mathbf{n}_{k}}{m}$
where $(\mathbf{n}_{k})_{i}=\delta _{ki}$. Hence, \ expanding to the lowest
non-trivial order of $1/m$%
\begin{eqnarray*}
&<&p_{k}>=<p_{k}>_{0}\left( 1+\frac{1}{m}\frac{\partial _{k}\pi }{\pi }%
\right) +O(\frac{1}{m^{2}}) \\
&<&(\Delta p_{k})^{2}>=<(\Delta p_{k})^{2}>_{0}\left( 1+\frac{2}{m}\frac{%
\partial _{k}\pi }{\pi }\right) +\frac{<p_{k}>_{0}^{2}}{m^{2}}\left( \frac{3%
}{2}\frac{\partial _{k}^{2}\pi }{\pi }-(\frac{\partial _{k}\pi }{\pi }%
)^{2}\right) +O(\frac{1}{m^{3}})
\end{eqnarray*}%
Thus, we recognize that the arbitrariness of the probability prior induces 
\emph{multiplicative noise} in the assigned probabilities, and affects their
variances both by rescaling and shifting. It is also worthwhile noticing
that the only instance of assigning zero probability would be due to the
choice of the prior; measurements alone, no matter how numerous, cannot
force us to assign strictly vanishing probabilities.

In the other extreme - no ($m=0$) data available - the probability
assignment derives through (eqn.2) strictly from the prior:%
\begin{equation*}
<p_{k}>=\mathcal{N}^{-1}\int d^{n}p\text{ }p_{k}\pi (\mathbf{p})\delta
(\Sigma _{i=1}^{n}p_{i}-1)=\overline{p_{k}}
\end{equation*}%
For one performed measurement ($m=1$) that produced a result in bin $i$ 
\begin{equation*}
<p_{k}>=\mathcal{N}^{-1}\int d^{n}p\text{ }p_{i}p_{k}\pi (\mathbf{p})\delta
(\Sigma _{i=1}^{n}p_{i}-1)=\frac{\overline{p_{k}p_{i}}}{\overline{p_{i}}}
\end{equation*}%
and analogously for higher values of $m$. Probabilities are most useful when
little or no data is available, and it is seen that such "ignorance"
probability assignments for measurable quantities are, not surprisingly,
entirely determined by the choice of the prior $\pi (p)$.

An interesting result is obtained when we go to the continuum limit $%
n\rightarrow \infty $. With $p_{k}=\int_{D_{k}}dxp(x)=\int_{x_{k}}^{x_{k}+%
\Delta x_{k}}dxp(x)=\Delta x_{k}p(x_{k})+\frac{1}{2}\Delta
x_{k}^{2}p^{\prime }(x_{k})+\cdots $ , the usual identification $%
p_{k}=\Delta x_{k}p(x_{k})$\ for $\Delta x_{k}\rightarrow 0$ only makes
sense when the probability density $p(x)$ is everywhere differentiable in $%
[a,b]$. In order to avoid handling ugly continual integrals, we perform the $%
n\rightarrow \infty $ limit at the stage where, with $\mu _{k}\equiv \frac{%
\pi (\varsigma _{k}^{\prime })}{\pi (\mathbf{\varsigma })}$ and $\sigma
(x_{k})\equiv \lim_{n\rightarrow \infty }$\ $\frac{1}{n\Delta x_{k}}$, 
\begin{equation*}
<p(x_{k})>=\mu (x_{k})\lim_{n\rightarrow \infty }\frac{<p_{k}>_{0}}{\Delta
x_{k}}=\mu (x_{k})\lim_{n\rightarrow \infty }\left[ \frac{m}{m+n}f(x_{k})+%
\frac{n}{m+n}\sigma (x_{k})\right]
\end{equation*}%
We observe that, for any finite amount of data ($m<\infty $) the assigned
probability density $<p(x)>=\mu (x)\sigma (x)$ depends on the metrics $%
\sigma $ and the prior but not on the data, while for $m=\infty $ the result
depends on the order in which the limits are taken. Only for $m\rightarrow
\infty $ before $n\rightarrow \infty $ is the result proportional to the
"sample frequency" density $f(x)$.

\emph{To summarize}, in order to relate probabilities (opinions) to the real
world (sample frequencies), we need the help of the Bayes theorem where a
probability prior enters the game. Hence, even when a lot of data is
available, the probability assignments are not unambiguous - the
arbitrariness of the prior manifests itself as a multiplicative noise in the
probabilities and in their variances. When little or no data is available
the assignments derive directly from the chosen prior. Let us also emphasize
an important lesson from the above: the widely held opinion that a
probability distribution represents a "state of knowledge" is wrong. It is
rather the sample frequencies, coming from observations, which constitute
"knowledge". Probabilities are necessarily inferred, and thus represent only
a "state of belief". The importance of this subtle distinction will become
apparent in what follows.

\section{Assigning Probabilities}

The most intellectually appealing scheme for assigning probabilities, in our
opinion, was put forward by Jaynes around the middle of the last century,
under the name "Maximum Entropy" (MaxEnt) principle. It is very difficult
for a rational person to argue with its general formulation, which simply
calls for inferential coherence by prescribing the assignment of the \emph{%
least committed} probability distribution consistent with \emph{all available%
} information. However, opinions rapidly start to diverge when it comes to
specifying how exactly the "least committed" distribution is defined and
what exactly constitutes "available information". On the first point, Jaynes
itself maintained that the "least committed" distribution is the one with
maximal Shannon entropy. His, and many others, affinity to Shannon's entropy
was based on a number of appealing properties it possesses. During the years
a tremendous amount of effort was invested into trying to prove that it is
"the one and only" reasonable criterion to use. Eventually, however, two
things were, or should have been, understood: 1) The Shannon's entropy is
but a particular instance of a larger class of equally reasonable Renyi's
entropies; and 2) The use of Jaynes procedure as a probability \emph{%
assignment} rule is untenable, so it gradually evolved into probability 
\emph{updating} rule - leaving us where we started, with the necessity of
assigning an ignorance prior. On the second point, the available information
is most often presented as a number of prescribed expectation values. Jaynes
himself was aware of the conflict between the expectations being
characteristics of probability distributions, and as such, essentially
opinions, and actual information obtained by measurements, but he took the
position that the available information entered in the form of constraint(s)
on the probability distribution even if \ "It might ... be only the guess of
an idiot" [3]. Before we embark on the ambitious task of trying to clarify
these points, let us briefly address the question of "once assigned, how can
probabilities be used?".

\subsubsection{What Use are Opinions?}

Probabilities being subjective, it is not immediately obvious how practical
use can be made of them. In statistical sense, probabilities are the best
estimators of sample frequencies, and this is about the only guiding
principle for their use. Hence, it appears that plugging probabilities in
place of sample frequencies in various statistical estimators would allow us
to infer \emph{predictions} about the results of measurements not yet
performed. Such statistical estimators are the Kolmogorov-Nagumo averages
[4], defined as $<A>_{\phi }=\phi ^{-1}\left( \sum p_{i}\phi (A_{i})\right) $
where $\phi (x)$ is continuous and strictly monotonic function, $A$ is an
observable, and $\ A_{i}$ is the value of $A$ corresponding to bin $i$.
Different functions $\phi $ in general produce different values of $%
<A>_{\phi }$. When measuring physical observables, we can use rulers in
different units and origin of the scale. Without an appropriate behaviour of
the predictions for the results of measurements upon rescaling and shifts
they would be useless. Therefore, an important requirement to be imposed on
an useful estimator is that $<\alpha A+\beta >_{\phi }=\alpha <A>_{\phi
}+\beta $, where $\alpha $ and $\beta $ are arbitrary constants. It is an
elementary exercise to show that this forces $\phi (x)=x$ and thus singles
out $<A>=\sum_{i}p_{i}A_{i}$ as the rule for predicting the result of a
measurement of $A$ given the probabilities $\{p\}\footnote{%
One might be tempted to argue in favor of the most probable value instead,
but one immediate indication that this is not a good rule is that it cannot
produce any prediction for uniform probabilities.}$. The result of an actual
measurement will most likely differ from the prediction, yet this is still
the best we can do with a probability assignment $\{p\}$.

\subsubsection{The Constraint Rule}

Let us first try to make the MaxEnt principle formulation more explicit in
its "using all avilable information" part. The physical problem under
consideration can be envisioned as the one of studying a set of observables
of a system, which we will refer to as "the primary observables". This could
be, e.g. the three coordinates $\mathbf{x}$ of a material point etc. We seek
to assign a probability distribution $p(\mathbf{x})$ for these observables,
which would allow us to a) Predict the results of future measurements of
these observables as their expectations $<\mathbf{x}>=\int dx\mathbf{x}p(%
\mathbf{x})$, which is of primary interest, and b) Predicting the result of
future measurements of any additional obervable $Q(\mathbf{x})$ \ as $%
<Q>=\int dxQ(\mathbf{x)}p(\mathbf{x}),$ which is of secondary interest. In
doing this, we are generally ignorant, except possibly for the results $%
\{a\} $ of previous measurements of some $m$ observables $A_{r},r=1,2,\cdots
,m$. Then the constraint rule of the MaxEnt principle can be regarded as a
requirement that the asigned probability distribution correctly "predicts"
the results of the already performed measurements as $a_{r}=\int dxA_{r}(%
\mathbf{x)}p(\mathbf{x}),r=1,2,\cdots ,m$. In other words, the constraint
rule simply forces the probability assignment, which is to be used to
predict the results of future measurements, to be consistent with the
results of measurements already performed. Let us stress that what is
involved here are single measurements and their results $\{a\},$ and not
multiple measurements from which the $a-s$ are obtained as sample averages,
as is too often implied in the context of the MaxEnt. Indeed, if the results
of, say, 10 measurements of, e.g., $A_{1}$ were known as $%
a_{1}(i),i=1,2,\cdots ,10$ and $a_{1}$ was taken as $\overline{a_{1}}=\frac{1%
}{10}\sum_{i=1}^{10}a_{i}(10)$ to be used as a constraint, this would be in
a blatant violation of the "using all available information" principle,
since the set of measured values of $A_{1}$ clearly contains information
also about $a_{1}^{\prime }$s variance: $\overline{\Delta a_{1}^{2}}=\frac{1%
}{9}\sum_{i=1}^{10}[a_{1}(i)-\overline{a_{1}}]^{2}$ and, similarly, for its
higher moments as well.

\subsubsection{The Expectation as (sort of) a Parameter}

Before we embark on the problem of assigning probabilities, we need to
shortly discuss the parameterization of our probability distributions in
terms of the expectations of their primary observables. For simplicity we
will assume one primary parameter $x$, the case with multiple such
parameters being a straight-forward generalization. In fact, we don't need
to consider a full-fledged parameterization in which the value of the
parameter is equal to the expectation of $x$, but just one that would allow
us to \emph{independently} vary the expectation of $x$. Thus, we are
interested in a parametrization $\ p(x;x_{e})$ such that, for any $%
|\varepsilon |<<1$, we have $\int dxxp(x;x_{e}+\varepsilon )=<x>+\varepsilon
+O(\varepsilon ^{3})$ while the normalization of the probability
distribution as well as all other cumulants $C_{n}(x)$ of $x$ are preserved%
\begin{equation*}
\frac{\partial }{\partial x_{e}}\int dxp(x;x_{e})=\frac{\partial }{\partial
x_{e}}\int dxC_{n}(x)p(x;x_{e})=0\text{ \ \ \ }n=2,3,\cdots
\end{equation*}%
We formulate the following \textbf{Conjecture}\footnote{%
The space restrictions do not allow us to formulate this as a theorem here.}%
: A parameterization with the above properties is only possible if the
probability distribution fulfills certain conditions at the border of its
domain, and in this case it is given by $p(x;x_{e})=p(x+x_{e})$. Obviously,
with such a parameterization we always have $\frac{\partial p(x;x_{e})}{%
\partial x_{e}}=\frac{\partial p(x;x_{e})}{\partial x}$, which is the
property we are mainly interested in. Establishing this, we finally can
address the "most uncommitted" element of the general MaxEnt principle.

\subsubsection{Assigning Robust Probabilities}

We have shown above that probability assignments based on observations have
inherent indeterminacy due to the arbitrariness of the probability prior.
Therefore, a natural question to ask is whether an assignment exsists that
is, in some sense, robust against variations of the prior. As already
demonstrated, the latter cause multiplicative noise in the probabilities.
Hence we try to formulate a robustness requirement in tems of a probability
distance of the Ali-Silvey type $D(p;p+\delta p)\underset{\{p\}}{\rightarrow 
}\min $, where $\delta p$ is the probability noise. As well known, for
normalization-preserving $\delta p(\mathbf{x})$ 
\begin{equation*}
D(p;p+\delta p)=\frac{\alpha }{2}\int dxp(\mathbf{x})\left( \frac{\delta p(%
\mathbf{x})}{p(\mathbf{x})}\right) ^{2}+O(\delta p^{3})
\end{equation*}%
where the constant coefficient $\alpha \sim 1$ depends on the particular
distance used. With a general multiplicative $\delta p(\mathbf{x}%
)=\varepsilon (\mathbf{x})p(\mathbf{x})$ the norm-preserving variation of
this with respect to $p(\mathbf{x})$ does not produce a solution. Hence, for
the most general probability noise our robustness requirement is not
selective enough to single out a particular distribution. However, upon some
reflection, we realize that not all possible perturbations in the
distribution are of equal importance: we are mainly interested in the
robustness of the probabilities with regard to the perturbations which would
have maximal effect on the primary observables, that is, choose the
multiplicative noise such that $\delta p(\mathbf{x})=\mathbf{\varepsilon (x)}%
p\mathbf{(x)}=\mathbf{\varepsilon }^{\prime }\mathbf{(x)}\cdot \frac{%
\partial p(\mathbf{x};\mathbf{x}_{e})}{\partial \mathbf{x}_{e}}$. With this
noise%
\begin{eqnarray*}
D(p;p+\delta p) &=&\frac{\alpha }{2}\int dxp^{-1}(\mathbf{x;\mathbf{x}%
_{e})\varepsilon }^{\prime }\mathbf{(x)}\cdot \frac{\partial p(\mathbf{x};%
\mathbf{x}_{e})}{\partial \mathbf{x}_{e}}\frac{\partial p(\mathbf{x};\mathbf{%
x}_{e})}{\partial \mathbf{x}_{e}}\cdot \mathbf{\varepsilon }^{\prime }%
\mathbf{(x)}+O(\varepsilon ^{\prime 3})\leq \\
&\leq &\frac{\alpha }{2}\int dx\mathbf{\varepsilon }^{\prime 2}\mathbf{(x)}%
\mathtt{Tr}I_{F}(\mathbf{x}_{e})
\end{eqnarray*}%
where $I_{F}(\mathbf{x}_{e})=\int dxp^{-1}(\mathbf{x;\mathbf{x}_{e})}$ $%
\frac{\partial p(\mathbf{x};\mathbf{x}_{e})}{\partial \mathbf{x}_{e}}\frac{%
\partial p(\mathbf{x};\mathbf{x}_{e})}{\partial \mathbf{x}_{e}}$ is the
Fisher information matrix with respect to $\mathbf{x}_{e}$ and the
inequality follows from its postivedefinitness. Hence, for an arbitrary
noise factor $\mathbf{\varepsilon }^{\prime }\mathbf{(x)}$ the tightest
bound on the distance results from the distribution with minimal trace of $%
I_{F}(\mathbf{x}_{e}).$ Using the interchangeability of the derivatives
derived above, we arrive at the final form of the robustness condition where 
$\mathbf{x}_{e}$ does not play a r\^{o}le any more and is therefore dropped%
\begin{equation*}
\int dxp^{-1}(\mathbf{x)}\left( \frac{\partial p(\mathbf{x})}{\partial 
\mathbf{x}}\right) ^{2}\rightarrow \min \text{ \ }\cap \text{\ \ }\int dxp(%
\mathbf{x})=1
\end{equation*}%
When results of measurements of some observables are known, the above
minimization is constrained such that the resulting probabilities reproduce
these observables. Is there any sense in which the so characterized
distribution could be considered "the least committed"? The Kramer-Rao
result for the most efficient estimator of $\mathbf{x}_{e}$ in the form $%
\mathtt{Tr}[cov^{-1}(\mathbf{x}_{e})]=\mathtt{Tr}I_{F}(\mathbf{x}_{e})$
indicates that in the situation where $p(\mathbf{x})$ is the one with
minimal trace of $I_{F}(\mathbf{x}_{e})$ an invariant measure of the
magnitude of the primary observables' covariance is maximal. This can be
formulated as "The distribution with minimal trace of the Fisher information
is the one for which the $\emph{most}$ $\emph{efficient}$ estimator of the
primary observables (whether it actually exists or not) has the \emph{worst
possible} performance". Thus the extremal property of $p(\mathbf{x})$ can
indeed\ be interpreted as the distribution being "the least committed" with
regard to the primary observables.

\section{Discussion}

Previously [5] we have derived the same condition (in the one-dimensional
case) for assigning uniformative probabilities from the requirement that
they be the least sensitive to coarse-graining. The rational for this
requirement was that coarse-graining decreased the "information content" --
if such a thing could be meaningfully defined -- and the distribution with
minimal information content to start with would be the one least affected by
it. The approach is, in a sense, complementary to Bernardo's reference
priors, where information is gained and the effect of this gain - maximized.
However, in contrast to Bernardo's, our result does not depend on which
particular distance is used to measure the sensitivity of the probabilities.
That the same assignment rule would result from the present, quite
different, considerations may bear some yet unidentified significance.
Fisher information-like constructs appear almost universally in physics [6]
and one cannot help but wonder to what extent physics laws could be
explained as information processing rules, and answer Toffoli's question
"Where does Nature shop for its Lagrangians".

Of significant interest is also whether/how the same considerations apply to
probabilities on discrete domains. In physics, discrete domains are usually
obtained by coarse-graining of continuous ones, and thus are "loaded" with
properties inherited from the topology and the metrics of the original
continuum. Such remnants could be, for example, various nearest, second
nearest etc. neighbour hierarchies. Choosing the simplest case of a
coarse-grained segment of the real line, the disrete domain is $%
\{1,2,...,n\} $ and the relevant observable is $<i>=\mathtt{nint}%
(\sum_{i=1}^{n}ip_{i}),$ where the function returns the nearest integer to
its argument. It can be conjectured as in the continuous case that the only
possible way to perturb the probabilities while best preserving their
normalization and higher cumulants of $i$, again subject to certain
conditions on $p_{1}$ and $p_{n}$, is equivalent to successive application
of $p_{i}^{\prime }=p_{i}-\varepsilon p_{i},$ $p_{i+1}^{\prime
}=p_{i+1}+\varepsilon p_{i}$, where the multiplicativity of the noise is
explicitely used. Then $D(p;p^{\prime })=\frac{\alpha }{2}\varepsilon
^{2}\sum_{i=1}^{n}p_{i}^{2}\left( \frac{1}{p_{i}}+\frac{1}{p_{i+1}}\right)
+O(\varepsilon ^{3})=\frac{\alpha }{2}\varepsilon ^{2}\left(
1+\sum_{i=1}^{n}p_{i}\frac{p_{i}}{p_{i+1}}\right) +O(\varepsilon ^{3})$. The
maximal robustness with respect to $<i>$ is achieved for a distribution for
which the distance is minimal, hence%
\begin{equation*}
\sum_{i=1}^{n}p_{i}\frac{p_{i}}{p_{i+1}}\rightarrow \min \text{ \ }\cap 
\text{ \ \ }\sum_{i=1}^{n}p_{i}=1
\end{equation*}%
Here the role of the Fisher information is played by the Renyi's distance of
order 2.

\paragraph{Acknowledgment}

This work was partially supported by the DTRA grant HDTRA1-11-1-0035.

\paragraph{References}

\begin{enumerate}
\item J. Tersoff, D. Bayer, Phys. Rev. Lett. 50, \#8 (1983) p.553

\item S. Sykora, \textit{J. Stat. Phys.} \textbf{11,} \#1 (1974) p.17

\item E.T.Jaynes, MaxEnt Formalism Conference, MIT, May 2-4, 1978

\item A. Kolmogorov, \textit{Atti della R. Accademia Nazionale dei Lincei} 
\textbf{12}, (1930) p.388; M. Nagumo, \textit{Japan. Journ. Math}. \textbf{7}%
, (1930) p,71

\item V.I.Dimitrov, \textit{AIP Conference Proceedings} Vol.\textbf{\ 954}
Issue 1, (2007) p.143

\item B. Roy Frieden, "\textit{Physics from Fisher Information}", Cambridge
Univ. Press, 1999
\end{enumerate}

\end{document}